\documentclass[twocolumn,superscriptaddress,showpacs,aps,prl,floatfix]{revtex4}
\usepackage[latin9]{inputenc}
\usepackage{textcomp}
\usepackage{amsmath}
\usepackage{amssymb}
\usepackage{graphicx}
\usepackage{esint}
\usepackage{bm}
\usepackage{comment}

\usepackage{color,ulem}

\makeatother
\newcommand{\bea}{\begin{eqnarray}}
\newcommand{\eea}{\end{eqnarray}}

\def\beq{\begin{equation}}
\def\eeq{\end{equation}}

\begin{document}

\title{Restoring the SU(4) Kondo regime in a double quantum dot system}
\author{L. Tosi}
\affiliation{Centro At\'{o}mico Bariloche and Instituto Balseiro, Comisi\'{o}n Nacional
de Energ\'{\i}a At\'{o}mica, 8400 Bariloche, Argentina}
\author{P. Roura-Bas}
\affiliation{Dpto de F\'{\i}sica, Centro At\'{o}mico Constituyentes, Comisi\'{o}n
Nacional de Energ\'{\i}a At\'{o}mica, Buenos Aires, Argentina}
\author{A. A. Aligia}
\affiliation{Centro At\'{o}mico Bariloche and Instituto Balseiro, Comisi\'{o}n Nacional
de Energ\'{\i}a At\'{o}mica, 8400 Bariloche, Argentina}
\date{\today }

\begin{abstract}
We calculate the spectral density and occupations of a system of two capacitively
coupled quantum dots, each one connected to its own pair of conducting
leads, in a regime of parameters in which the total coupling to the leads for each dot $\Gamma_i$
are different. 
The system has been used recently to perform pseudospin spectroscopy
by controlling independently the voltages of the four leads. 
For an odd number of electrons in the system, $\Gamma_1=\Gamma_2$, equal dot levels $E_1=E_2$ 
and sufficiently large interdot repulsion
$U_{12}$ the system lies in the SU(4) symmetric point of spin and pseudospin degeneracy in the Kondo regime.  
In the more realistic case $\Gamma_1 \neq \Gamma_2$, pseudospin degeneracy is broken 
and the symmetry is reduced to SU(2). Nevertheless we find that the essential features 
of the SU(4) symmetric case are recovered by appropriately tuning the level difference
$\delta=E_2-E_1$. The system behaves as an SU(4) Kondo one at low energies. 
Our results are relevant for experiments which look for signatures of 
SU(4) symmetry in the Kondo regime of similar systems.
\end{abstract}

\maketitle


\section{Introduction} 
\label{intro}

The Kondo effect,
is one of the most studied phenomena in strongly correlated
condensed matter systems \cite{hew} and continues to be a subject of great interest. 
The effect is characterized by the emergence of a
many-body singlet ground state formed by the impurity spin and the
conduction electrons in the Fermi sea.  
The binding energy of this singlet is of the order of the characteristic Kondo temperature 
$T_K$ below which the
effects of the ``screening'' of the impurity spin manifest in different physical properties.
The first observed manifestation of the Kondo effect was the logarithmic increase of
the resistivity as the temperature is decreased in systems of magnetic impurities in metals.\cite{kondo}
The same type of behavior is present in systems with orbital degeneracy
but no spin degeneracy.\cite{matias}
In the last decades, the research has moved
to nanoscopic systems with    
semiconducting \cite{gold1,cro,gold2,wiel} or molecular \cite{parks,serge,parks2}
quantum dots (QDs), 
with a single ``impurity'', in which different parameters like on-site energy and
hybridization of the impurity with the conduction electrons can be controlled very well.

In the last years there has been research on Kondo systems in which in addition to
the spin degeneracy, there is also degeneracy in other ``orbital'' degree 
of freedom such that the complete symmetry of the system is very high, corresponding to the 
SU(4) Lie group 
\cite{borda,zar,karyn,desint,jari,choi,lim,ander,lipi,buss,fcm,grove,tetta,see,mina,lobos,buss2,keller,oks,nishi,fili}.
Some examples are
quantum dots in carbon nanotubes \cite{jari,choi,lim,ander,lipi,buss,fcm,grove},
silicon nanowires \cite{tetta}, and organic molecules deposited on Au(111) 
\cite{mina,lobos}.

More recently a double QD with strong interdot capacitive coupling, and each QD
tunnel-coupled to its own pair of leads has been experimentally 
studied.\cite{keller,ama}
The occupation of one QD or the other plays the role of the
orbital degree of freedom, and behaves as a pseudospin. These occupations, the tunneling 
matrix elements (coupling to the leads) and the
voltages at the four leads can be controlled independently. 
The system was proposed by B\"{u}sser {\it et al.} \cite{buss2} to control
spin-polarized currents and is being subject of intense theoretical research
\cite{keller,oks,nishi,fili,arx1,bao}. Comparing experiment with numerical-renormalization-group (NRG)
calculation, Keller {\it et al.} found
evidence of SU(4) Kondo behavior.\cite{keller} However, using renormalized perturbation 
theory (RPT) with parameters obtained from NRG, Nishikawa {\it et al.} conclude that the 
experimental system is not in the SU(4) regime, particularly because of the relatively small
value of the interdot repulsion $U_{12}$ in comparison with other parameters.\cite{nishi}
In addition, it is in principle a difficult to reach the SU(4) condition $\Gamma _1=\Gamma _2$, where 
$\Gamma _i=\Gamma _{S i}+\Gamma _{D i}$ and 
$\Gamma _{\nu i}$ is the coupling of the source ($\nu=S$) or drain ($\nu=D$) lead
with dot $i$. The total coupling of dot $i$, $\Gamma _i$, corresponds to the line width of the 
local spectral density of dot $i$ in absence of Coulomb repulsion and is inferred from
experiment with the aid of theory.\cite{keller} Instead, the energy at each dot $E_i$
is easier to control directly by the different applied voltages as described in the
supplementary material of Ref. \onlinecite{ama}.

The purpose of the present work is to study to what extent the loss of SU(4) symmetry caused
by unequal couplings $\Gamma _1 \neq \Gamma _2$ can be restored by tuning the energy
difference $\delta=E_2-E_1$,  in a regime of parameters in which intrasite $U_i$ and intersite 
$U_{12}$ repulsions are much larger than the $\Gamma _i$. 
This is related with the concept of {\it emergent symmetry}
i.e., the fact that new symmetries not realized in the Hamiltonian
describing the system can emerge at low energies.\cite{eme}

We use the non-crossing approximation (NCA) which has been applied to similar systems
and has the advantage of been easily extensible to the non-equilibrium case of finite bias 
voltages.\cite{tetta,oks,nca,nca2,benz}
In fact, we have previously studied the conductance of the system of two capacitively coupled quantum 
dots in the general case of different finite bias voltages $V_i$ applied to each dot.\cite{oks} 
We have discussed 
the conditions
to observe an SU(4) $\rightarrow$ SU(2) crossover under an applied pseudo-magnetic field $\delta$, 
and non-trivial crossed effects 
of changes in the conductance through one QD as a voltage is applied to the other. 
Recently the general non-equilibrium case has been studied using equations of motion.\cite{bao}
However, as in 
most previous theoretical studies of the system, $\Gamma _1 = \Gamma _2$ was assumed.
An alternative to study the non-equilibrium case for small $V_i$ might be
to use RPT,\cite{nishi,hrpt,ogu,scali} but its extension to the two-dot case 
and finite $V_i$ seems difficult because of the presence of many parameters.\cite{nishi}

In this paper we calculate the spectral densities $\rho_i$ of each dot. They can be addressed experimentally
in a situation with very asymmetric coupling to the source and drain leads for each dot $i$, changing only the 
voltage to the less coupled lead. 
In fact a ratio 
$\Gamma_{Si}/\Gamma_{Di}=10$
or 0.1 is enough for the differential conductance $dI/dV$ to represent accurately $\rho_i$,\cite{oks} 
and a ratio 12 has been used in some experiments.\cite{ama}
We show that the main effect of different total couplings of both dots $\Gamma _1 \neq \Gamma _2$
is to introduce an effective pseudo-Zeeman splitting $\delta_{\rm eff}$. This can be understood
by a straightforward generalization of the scaling treatment of Haldane for the simplest impurity
Anderson model (corresponding to the one-dot case).\cite{hald} This $\delta_{\rm eff}$ 
can be compensated tuning the gate voltages so that $\delta=E_2-E_1=-\delta_{\rm eff}$ leading to an 
SU(4) behavior at low energies. 

The paper is organized as follows. The model is explained in Section \ref{model}. In Section
\ref{su42} we explain the main differences in the spectral densities in the regimes where the SU(4) 
or SU(2) symmetry and the ``transition'' between them. Section \ref{dgam} describes the effect of different
total couplings to the leads $\Gamma _{2} \neq \Gamma _{1}$. In Section \ref{restoring} we 
describe how tuning the energy levels can compensate the effect of different couplings
in a restricted energy range. Section \ref{summ} contains a summary and a discussion.

\section{Model}
\label{model}

The system is described by an Anderson model which contains as localized configurations
a singlet $|s\rangle $ with an even number of particles in each dot and two 
spin doublets $|i\sigma\rangle $ ($i=1$ or 2) with one additional electron (or hole) in
QD $i$. There are four conduction bands which correspond to separate source and drain leads 
for each dot. The Hamiltonian is 
\begin{eqnarray}
H &=&E_{s}|s\rangle \langle s|+\sum_{i\sigma }E_{i}|i\sigma \rangle \langle
i\sigma |+\sum_{i \nu k \sigma }\epsilon _{\nu k}c_{\nu k i\sigma }^{\dagger
}c_{\nu ki\sigma }  \notag \\
&&+\sum_{i \nu k\sigma }(V_{i}^{\nu }|i\sigma \rangle \langle s|c_{\nu
ki\sigma }+\mathrm{H.c}.),  \label{ham}
\end{eqnarray}%
where $c_{\nu ki\sigma }^{\dagger }$ create conduction states at the source 
($\nu =S$) or drain ($\nu =D$) lead, and $V_{i}^{\nu }$ is the hopping
between the lead $\nu $ and dot $i$, assumed independent of $k$.  Since charge
configurations with two particles are excluded, the model assumes infinite
on-site repulsions $U_{i}$ and interdot repulsion $U_{12}$.

The tunnel couplings of each QD to the leads are $\Gamma _{\nu i}=2\pi
\sum_{k}|V_{i}^{\nu }|^{2}\delta (\omega -\epsilon _{\nu k})$,  and we take
the unit of energy $\Gamma _{1}=\Gamma _{S1}+\Gamma _{D1}=1$ unless
otherwise stated and (without loss of generality) $\Gamma _{2} \leq \Gamma _{1}$  

\section{The SU(4) $\rightarrow$ SU(2) crossover}
\label{su42}

For future comparison, in this section we review the effect of Zeeman or 
pseudo-Zeeman splitting
on the SU(4) Anderson model,\cite{desint,fcm,tetta} and the effects of temperature.

\begin{figure}[h]
\begin{center}
\vspace{0.5cm}
\includegraphics[width=7cm]{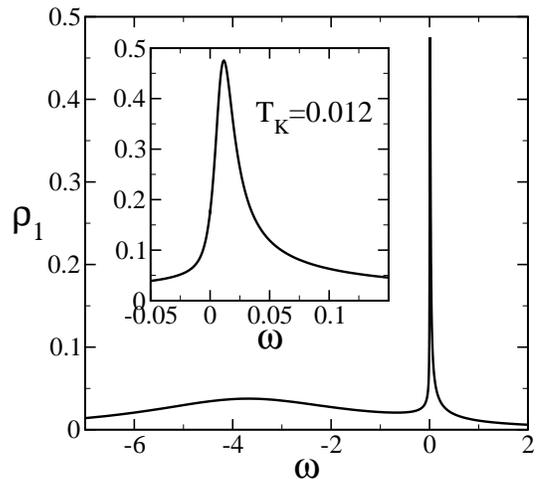}
\caption{Density of states at each dot as a function of energy 
at low temperature $T=5.10^{-3}=0.42\ T_K$ in the SU(4) case $E_1=-4$ 
and $\Gamma_1=\Gamma_2=1$. The inset is a detail close to the Fermi level.}  
\label{Fig:SU4}
\end{center}
\end{figure}

In Fig. \ref{Fig:SU4} we show the spectral density of states per spin at each dot $\rho_1=\rho_2$ 
(the spin subscript is dropped)
in an SU(4)-symmetric case $E_1=E_2=-4$ and $\Gamma_{1}=\Gamma_{2}=1$.
We take the half band width $D=10$ for all calculations presented here.
The density $\rho_i$ corresponds to the operators $|i\sigma\rangle \langle 0 |$ 
(see Ref. \onlinecite{benz} for details). 
The density of states shows two peaks. The charge-transfer one is broad of 
half width at half maximum near $\Gamma_1$ (two times that in the non-interacting case). The Kondo peak near
the Fermi level has a half width at half maximum of the order of the Kondo temperature $T_K$.  
As a consequence of the increase of degeneracy, the Kondo effect is stronger than for the usual SU(2) case. 
We remind the reader that the Kondo temperature for the infinite-$U$ SU(N) Anderson model is 
$T_K \approx D {\rm exp} [\pi E_1/(N \Delta)]$ where $D$ is half the band width 
and $\Delta=\Gamma_1/2$.\cite{hew} The NCA reproduces correctly this result.\cite{bickers}

From half the width of the spectral density we obtain $T_K^{\rm SU(4)}=0.012$.
In comparison with the SU(2) case with the same $T_K$, the Kondo resonance is displaced to 
higher energies and the maximum is
clearly above the Fermi energy which we set as the origin of energies ($\epsilon_F=0$).
In fact, the SU(4) case is characterized by a high derivative of $\rho_i (\omega)$ at the Fermi level,
leading to a large thermoelectric power.\cite{see}

Although the symmetry is broken immediately when even a tiny pseudo-Zeeman splitting $\delta$ is introduced, 
the changes in physical quantities like conductances for each dot and occupations are not appreciable until 
$\delta$ becomes of the order of $T_K^{\rm SU(4)}$.\cite{desint} In particular, 
the Kondo temperature $T_K(\delta)$ obtained 
from the width of the Kondo peak displays initially a plateau and then decreases strongly for  
$\delta > T_K^{\rm SU(4)}$. We have obtained that our NCA results for $T_K(\delta)$ can be very well represented 
by a simple equation obtained from a variational wave function

\begin{equation}
T_{K}=\left\{ (D+\delta )D\exp \left[ \pi E_{1}/(4\Delta )\right] +\delta
^{2}/4\right\} ^{1/2}-\delta /2.   \label{tk}
\end{equation}
times a factor of the order of 1 (0.606 for the parameters used).\cite{fcm}

\begin{figure}[h]
\begin{center}
\includegraphics[width=7cm]{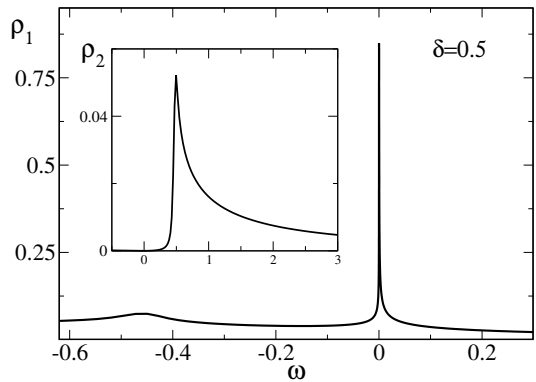}
\caption{Density of states of dot 1 (main figure) and dot 2 (inset) as a function of energy 
at low temperature $T=5 \times 10^{-5}$ for $E_1=-4$, $E_2=E_1+\delta$ with $\delta=0.5$ 
and $\Gamma_1=\Gamma_2=1$.}  
\label{Fig:delta}
\end{center}
\end{figure}

For $\delta > T_K^{\rm SU(4)}$ the changes in the spectral density at low temperatures are dramatic, as
shown in Fig. \ref{Fig:delta}.
The Kondo peak narrows [following Eq. (\ref{tk})], and shifts towards the Fermi energy. 
The density of the dot that corresponds 
to the lowest lying level (1 in our convention) develops a peak at energy near $-\delta$ while the other 
density ($\rho_2$) displays only a peak near 
$+\delta$ but no Kondo peak. 

\begin{figure}[h]
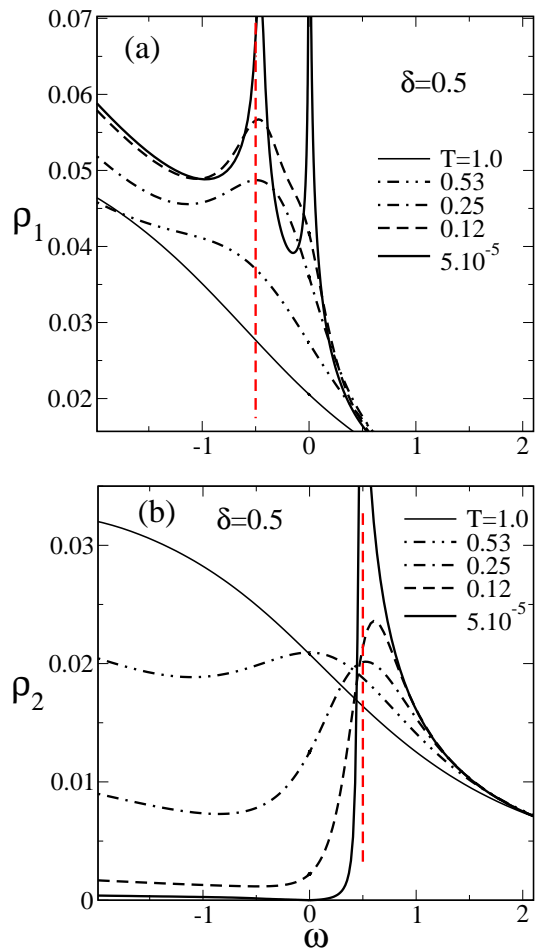

\begin{center}
\includegraphics[width=7cm]{rho_1_n.eps}\\
\includegraphics[width=7cm]{rho_2_n.eps}
\caption{Density of states of (a) dot 1 and (b) dot 2 as a function of energy for different
temperatures. Parameters as in Fig. \ref{Fig:delta}.}
\label{Fig:delta2}
\end{center}
\end{figure}

It is interesting to see the evolution of the side peaks at $\pm \delta$ with temperature. It is shown in 
Fig. \ref{Fig:delta2}. At high temperatures $T > \delta$ both spectral densities are similar. As 
the temperature is lowered below $\delta$ the side peaks start to develop. In addition, as the total occupation of
dot 1 $n_1=n_{1 \uparrow}+n_{1 \downarrow}$ increases and that of dot 2 ($n_2$) decreases, 
the charge transfer peak of dot 1 (2) increases 
(decreases). In the figure, due to the restricted energy range, only the tail of this peak is visible, 
but the above mentioned effect is clear. The changes in $n_i$ as a function of $\delta$ 
were studied before.\cite{desint,nishi} At temperatures below $T_K(\delta)=0.155$ ($T_K(0.5)=0.155$ in the figure),  
the Kondo peak develops in $\rho_1$. The width of both side peaks is of the order of $T_K(0)=T_K^{\rm SU(4)}$.

It is important to recall that in general, in the presence of both Zeeman and pseudo-Zeeman splitting, the spectral 
density at the Fermi level at zero temperature for dot $i$ and spin $\sigma$ is related to the corresponding 
occupation by the Friedel sum rule generalized for orbital degeneracy.\cite{yoshi} We assume that the $\Gamma_i$ and the
unperturbed densities of conduction states are independent of energy. Since $T_K(\delta)$ is always
much smaller than typical scales of variations of this parameters, this assumption is realistic. 
In this case, the Friedel sum rule simplifies to \cite{fcm,yoshi}

\begin{equation}
\rho _{i \sigma}(\epsilon _{F})=\frac{1}{\pi \Delta }\sin ^{2}(\pi n_{i \sigma}).
\label{fsr}
\end{equation}

At $\delta=0$ the four occupations $n_{i \sigma}$
are slightly below 1/4 (the total occupation is below 1 because of a finite small occupation of the 
singlet $|0 \rangle$).
For finite $\delta$ and high temperatures in comparison with $\delta$, also all $n_{i \sigma}$
are slightly below 1/4. As the temperature decreases below $\delta$, with $\delta$ large in comparison 
with $T_K(0)$  $n_{1 \sigma}$ increase towards 1/2 
while $n_{2 \sigma}$ decrease towards 0. The Friedel sum rule implies that at $T=0$, 
$\pi \Delta \rho _{i \sigma}$ is
slightly below 1/2 in the SU(4) case, while well inside the SU(2) regime,
$\pi \Delta \rho _{1 \sigma} \rightarrow 1$ (or slightly below) and $\pi \Delta \rho _{2 \sigma} \rightarrow 0$. 
The NCA has an error of the order of 15 \% in the Friedel sum rule, but the tendencies are well 
reproduced.\cite{fcm}

\section{Effect of different couplings for degenerate levels}
\label{dgam}

Starting from degenerate levels $E_1=E_2=E_d$, the main effect expected from different total hybridizations
$\Gamma _1 \neq \Gamma _2$ is to generate an effective pseudo-Zeeman splitting $\delta_{\rm eff}=E_2^*-E_1^*$,
where $E_i^*$ are renormalized energy levels. This can be seen generalizing the theory used by Haldane
based on poor man's scaling to find $E^*$ for the case of one level.\cite{hald} 
One proceeds integrating out the states near the top (with energy $+D$) and bottom (energy $-D$)
of the conduction band.
The localized state can be empty with energy $e_0$ or occupied with energy $e_{i \sigma}$. 
After renormalization, the energy necessary to add one localized particle is $E^*_{i \sigma}=e_{i \sigma}-e_0$.  
The renormalization is caused by the possible processes of destroying an electron in the localized 
level and creating it in the conduction band or vice versa. When being integrating out, 
each state near the bottom of the conduction band contributes to lowering the energy of the empty state 
$e_0$ by 
\begin{equation}
\sum_{i \sigma} \frac{V_{i \sigma}^2}{E_d+D}. \nonumber
\end{equation}
Similarly the states near the bottom of the conduction band lower the energy of the occupied state 
$e_{i \sigma}$ by 
\begin{equation}
\frac{V_{i \sigma}^2}{D-Ed}. \nonumber
\end{equation}
Scaling down to a cutoff $C$ one obtains
\begin{equation}
E^*_{i \sigma}=\frac{1}{2\pi}[\sum_{j \sigma^\prime}\Gamma_{j \sigma^\prime}-\Gamma_{i \sigma}]\ln \left(\frac{D}{C}\right).
\label{e_eff}
\end{equation}
In our case in which the couplings are independent of spin, this leads to an effective splitting
\begin{equation}
\delta_{\rm eff}=\frac{1}{2\pi}(\Gamma_2-\Gamma_1)\ln \left(\frac{D}{C}\right).
\label{d_eff}
\end{equation}

\begin{figure}[h]
\begin{center}
\includegraphics[width=8cm]{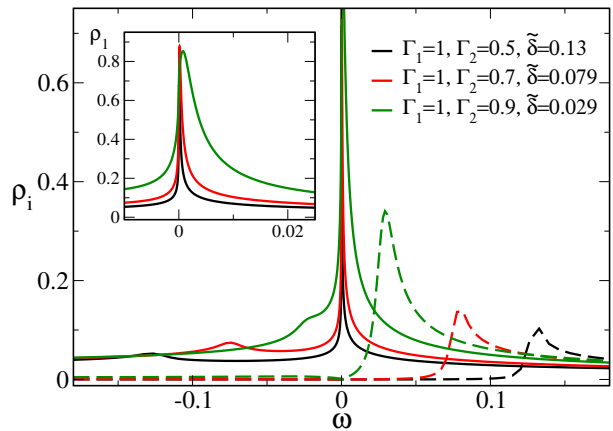}
\caption{(color online) Density of states of dot 1 (full lines) and dot 2 (dashed lines) as a function of frequency 
for low temperatures ($T=5.10^{-5}$), $E_1=E_2=-4$, $\Gamma_{1}=1$, and several values of $\Gamma_{2}$. 
The inset shows a detail near the Fermi energy.}
\label{Fig:difhyb}
\end{center}
\end{figure}

In Fig. \ref{Fig:difhyb} we display the spectral density of states for three cases with $\Gamma_1>\Gamma_2$.
A comparison with the results of the previous section indicates that the effects of different total coupling
to the leads for both dots are similar to those of a splitting of the energy levels. The Kondo peak at the 
Fermi energy narrows and displaces towards the Fermi energy and a side peak appears for each dot, at negative 
(positive) energies for the more (less) coupled dot.  
For the case $\Gamma_{2}=0.9$, only 10 \% less than $\Gamma_1$, the side peak in $\rho_1$ appears as a 
shoulder to the left of the Kondo peak rather than being well separated, because the effective splitting 
$\delta_{\rm eff} \approx 0.029$ is of the order of the Kondo temperature for $\Gamma_i=1$, 
$T_K^{\rm SU(4)}=0.012$.

\begin{figure}[t]
\begin{center}
\includegraphics[width=6cm]{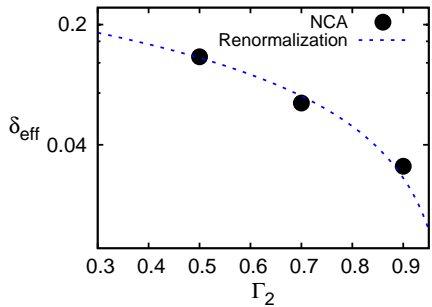}
\caption{Effective splitting as a function of $\Gamma_{2}$.
(color online) The dashed blue line is the result from Eq. (\ref{d_eff}) with $C=|E_d/2|$.
Other parameters as in Fig. \ref{Fig:difhyb}.}
\label{Fig:splitting}
\end{center}
\end{figure}

The positions of the side peaks allow us to infer the values of the effective splitting within the NCA.
They are 
listed inside Fig. \ref{Fig:difhyb} as $\tilde{\delta}$ and represented in Fig. \ref{Fig:splitting}
together with the result of Eq. (\ref{d_eff}) with a cutoff $C=|E_d/2|$. 
In his scaling calculation, Haldane used a cutoff of the order of $\Gamma_i$,\cite{hald} 
while in a recent detailed study of the prefactor of the Kondo temperature of the SU(4) 
case, Filipone {\it et al.} used $C=|E_d|/\alpha$ with $\alpha$ of the order of 1.\cite{fili}
We obtain a better agreement with the NCA results using the latter choice.
The good agreement between both approaches (in spite of the corresponding limitations of
each one) seems to confirm the physical picture of the main effect of different couplings.

\section{Restoring SU(4) symmetry}
\label{restoring}

After the results of the previous section, the question arises if introducing a real 
splitting $\delta=E_2-E_1$ such that it compensates the effect of different couplings
(so that $\delta+\delta_{\rm eff}=0$), the SU(4) symmetry can be restored in 
the low-energy properties tested by conductance measurements. Clearly the symmetry
remains broken at the Hamiltonian level, so that one cannot expect a higher symmetry at all 
energies. Therefore, we search for indications of a low-energy {\it emergent symmetry}.\cite{eme}

\begin{figure}[h!]
\begin{center}
\includegraphics[width=7cm]{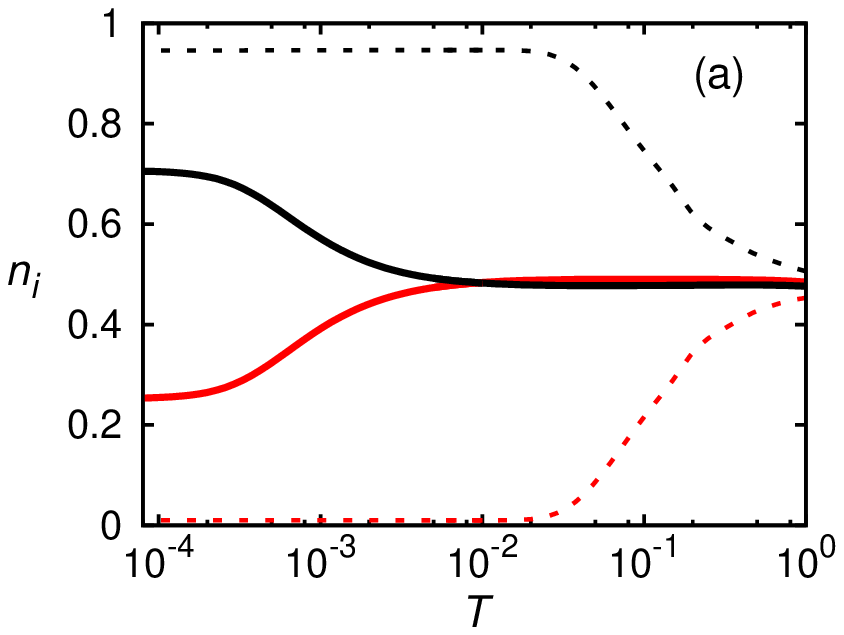}\\
\vspace{-0.5cm}
\includegraphics[width=7cm]{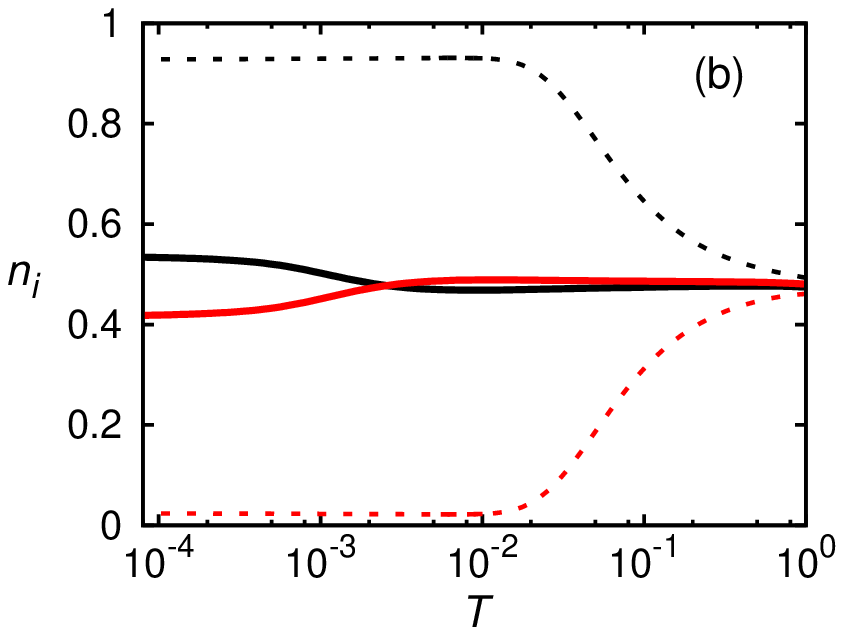}\\
\vspace{-0.5cm}
\includegraphics[width=7cm]{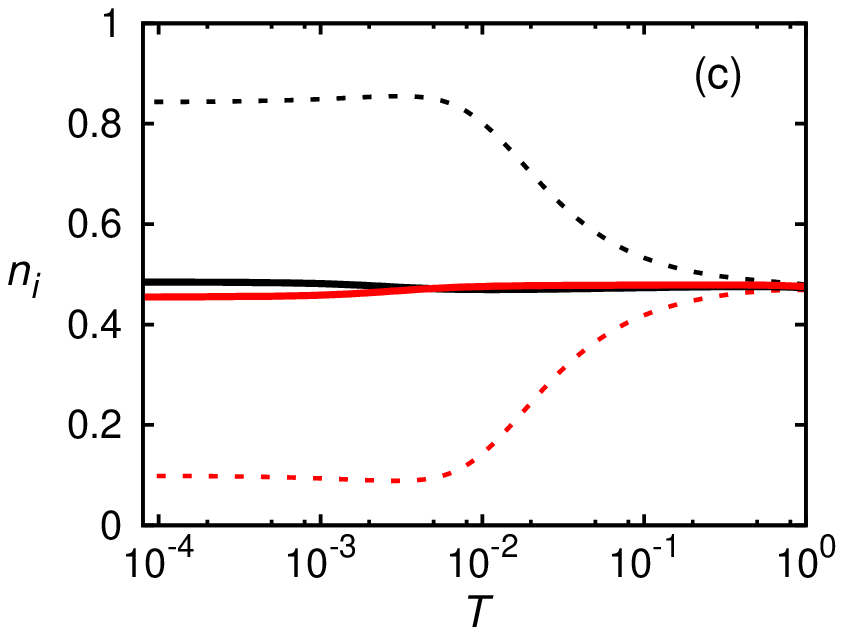}
\caption{(color online) Total occupations of dot 1 (black) and dot 2 (red) 
as a function of temperature for $E_1=-4$, $E_2=E_1+\delta$,   
$\Gamma_{1}=1$,  and (a) $\Gamma_{2}=0.5$, $\delta=-0.13$, (b) $\Gamma_{2}=0.7$, $\delta=-0.08$ and (c)
$\Gamma_{2}=0.9$, $\delta=-0.03$. The case $\delta=0$ (dashed lines) is shown for comparison.}
\label{Fig:restore}
\end{center}
\end{figure}

\begin{figure}[h!]
\begin{center}
\includegraphics[width=8cm]{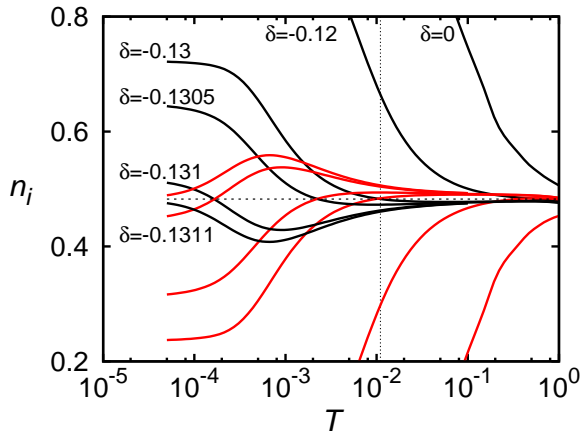}
\caption{(color online) Total occupations of dot 1 (black) and dot 2 (red) 
as a function of temperature for $\Gamma_{2}=0.5$, and several values of $\delta=0,\ -0.12,\ -0.13,\ -.1305,\ -0.131,\ -0.1311$. Other parameters as in Fig. \ref{Fig:restore}.}
\label{Fig:tunning05}
\end{center}
\end{figure}

In Fig. \ref{Fig:restore} we show the temperature dependence of the total occupations (adding both spins) 
at each dot for a splitting such that $\delta+\delta_{\rm eff} \approx 0$ according to the results of the 
previous section. At high temperatures, of the order of $\Gamma$, $n_1 \approx n_2$, although $n_2$ 
(the occupation of the less hybridized doublet lying at lower energy) 
is slightly larger. As the temperature is lowered by two order of magnitude, the situation is similar, 
although $n_2-n_1$ first increases slightly and then decreases. At temperatures below $T_K^{\rm SU(4)}$,
$n_2-n_1$ changes sign and increases in magnitude, signaling a complete loss of SU(4) symmetry for 
$T \rightarrow 0$. However, it is possible to tune $\delta$ so that the condition $n_2=n_1$ 
(implied by SU(4) symmetry) is satisfied at any given temperature. Conversely, for a given
$\delta$, $T$ can be varied so that $n_2=n_1$ at $T=T_{\rm occ}(\delta)$. In fact, the choice 
$\delta=-\delta_{\rm eff}$ with $\delta_{\rm eff}$ extracted from the position of the satellite
peaks is a good initial guess, but tuning $\delta$, $T_{\rm occ}(\delta)$ can be reduced by orders of magnitude, 
as shown in Fig. \ref{Fig:tunning05}. This tuning is very time consuming for our numerical procedure to solve
the selfconsistent set of NCA equations (for details see for example Ref. \onlinecite{benz}) because 
the procedure has to be repeated for several ``guessed'' values of $\delta$ near $\delta_c$,
where $\delta_c$ is defined by $T_{\rm occ}(\delta_c)=0$. In addition
the NCA cannot reach arbitrarily small temperatures. It is interesting to note 
that we find that $T_{\rm occ}(\delta)$ has a nearly exponential 
dependence near $\delta_c$. As $\delta$ is varied between -0.13 and -0.131, $T_{\rm occ}$ decreases from
$10^{-2}$ (of the order of $T_K^{\rm SU(4)}$) to $10^{-4}$.
Note that for sufficiently negative $\delta$ ($\delta<\delta_c$), 
$n_2$ remains larger than $n_1$ and there is no crossing point with $n_1=n_2$.

If $n_1=n_2$ at $T=0$ (for $\delta=\delta_c$), the Fridel sum rule Eq. (\ref{fsr}) 
implies also that the spectral densities at the Fermi energy are equal: $\rho_1(0)=\rho_2(0)$.
This is difficult to test in a conductance measurement, because the conductance
through each dot is proportional to the asymmetry factor 
$A_i=4 \Gamma_{Si} \Gamma_{Di}/(\Gamma_{Si}+\Gamma_{Di})^2$, and these factors are not
easy to be determined precisely.\cite{keller,ama}
However, as we have explained the line shape of the spectral densities are very different 
in the SU(4) and SU(2) regimes, not only because of the presence of the satellite peaks but 
also for the different shape of the Kondo peak which in turn implies for example a different
temperature dependence of the equilibrium conductances.\cite{lim,see,keller}

\begin{figure}[t!]
\begin{center}
\includegraphics[width=8cm]{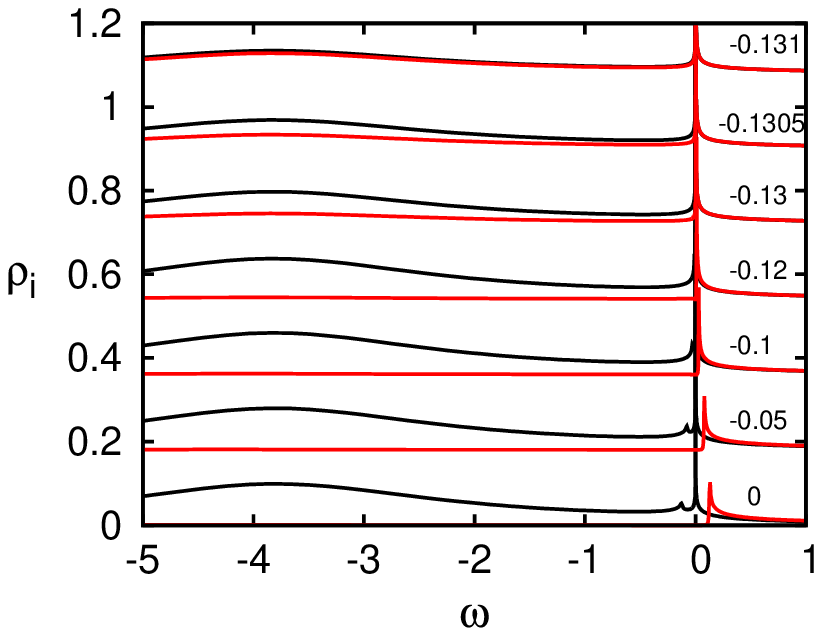}\\
\includegraphics[width=8cm]{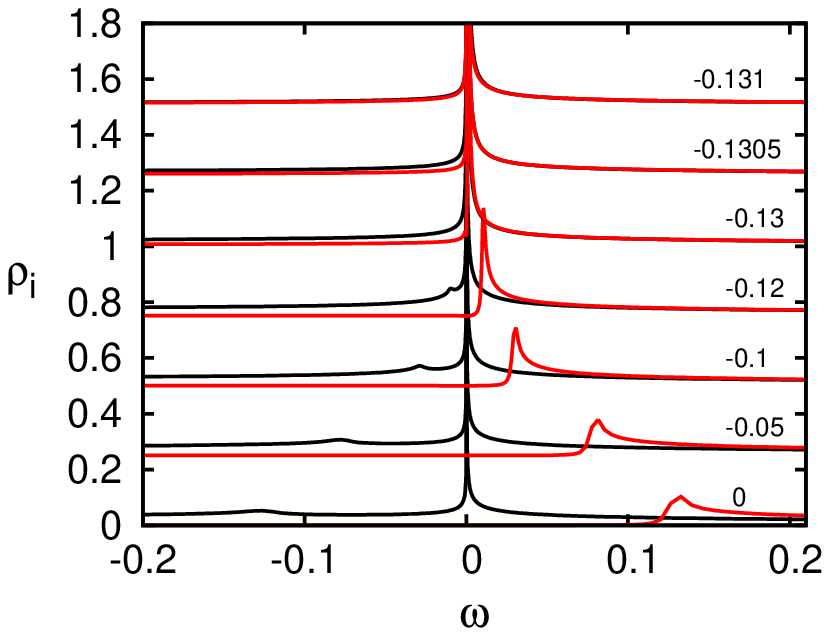}
\caption{(color online) Spectral density of dot 1 (black) and dot 2 (red) 
as a function of energy for several values of $-\delta:0,\,0.05,\,0.1,\,0.12,\,0.13,\,0.1305$ 
and $0.1309$. 
The upper figure corresponds to an extended energy range.
The temperature is $T\approx 5.10^{-5}$. 
Other parameters as in Fig. \ref{Fig:tunning05}.
The curves for $\delta \neq 0$ are shifted with a vertical offset for clarity.}
\label{Fig:density}
\end{center}
\end{figure}

In Fig. \ref{Fig:density} we show the evolution of the low-temperature densities of states 
with $\delta$, as the crossing point $n_1=n_2$ is approached lowering $\delta$ from 0. 
The first rather obvious change is that as $n_1$ decreases and $n_2$ increases, the 
weight of the corresponding charge-transfer peaks near $\omega=-4$ change roughly proportional
to $n_i$ until they become almost coincident. The changes near the Fermi energy are more subtle 
and they resemble the opposite of those reported in section \ref{su42}:
the side peaks move towards the Fermi energy, the Kondo resonance in $\rho_1(\omega)$ broadens and 
displaces partially to higher energies, a Kondo resonance appears in $\rho_2(\omega)$ and both
densities tend to merge.

\begin{figure}[t!]
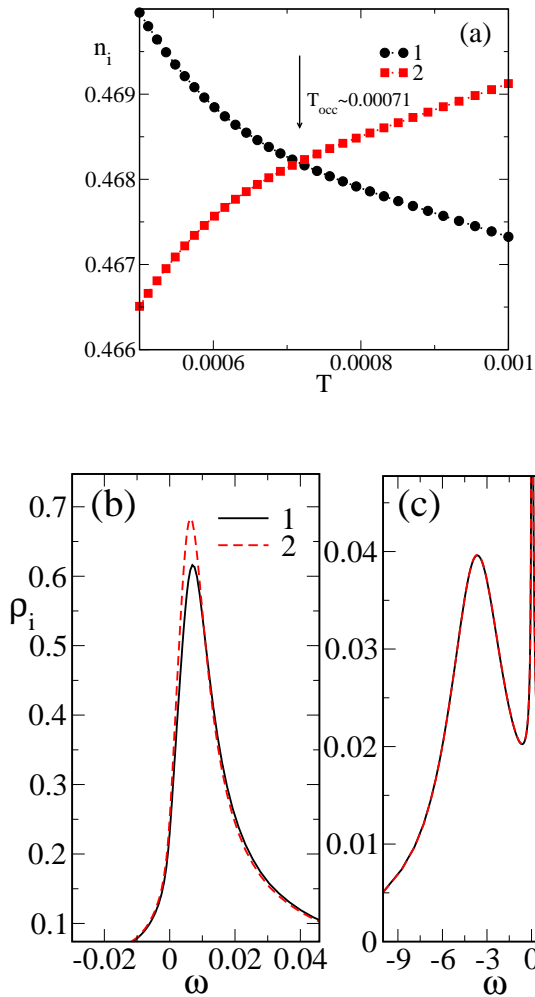

\begin{center}
\includegraphics[width=6cm]{compara-occ-2.eps}\\
\vspace{1.0cm}
\includegraphics[width=8cm]{compara-rho-Tocc.eps}
\caption{(color online) (a) Occupation as a function of temperature and (b), (c) spectral 
density at $T=T_{\rm occ}(\delta)$ as a function of energy of
dot 1 (black) and dot 2 (red),  for $\Gamma_1=1$ and $\Gamma_2=0.9$, $E_1=-4$, and $\delta=-0.031$.}
\label{Fig:compara}
\end{center}
\end{figure}

As explained at the beginning of this section, we do not expect that for any parameters, 
$\rho_1(\omega)$ and $\rho_2(\omega)$ coincide for all energies. 
In Fig. \ref{Fig:compara} we compare both densities 
at $T=T_{\rm occ}(\delta)$ for a 10 \% mismatch in the $\Gamma_i$. 
The charge-transfer peaks look identical, 
but the maxima of the Kondo resonances differ by about 10 \%, being higher for dot 2 (the lowest lying 
and less hybridized level). The magnitude of $\rho_i(0)$, the densities at the Fermi level are also 
slightly different being $\rho_2(0)>\rho_1(0)$ by near 10 \%. 
This might be an effect of finite temperature or of the inaccuracy
of the NCA to reproduce the Friedel sum rule Eq. (\ref{fsr}). In any case, the shape of both
densities are characteristic of the SU(4) regime and they are quite similar. 

\begin{figure}[h]
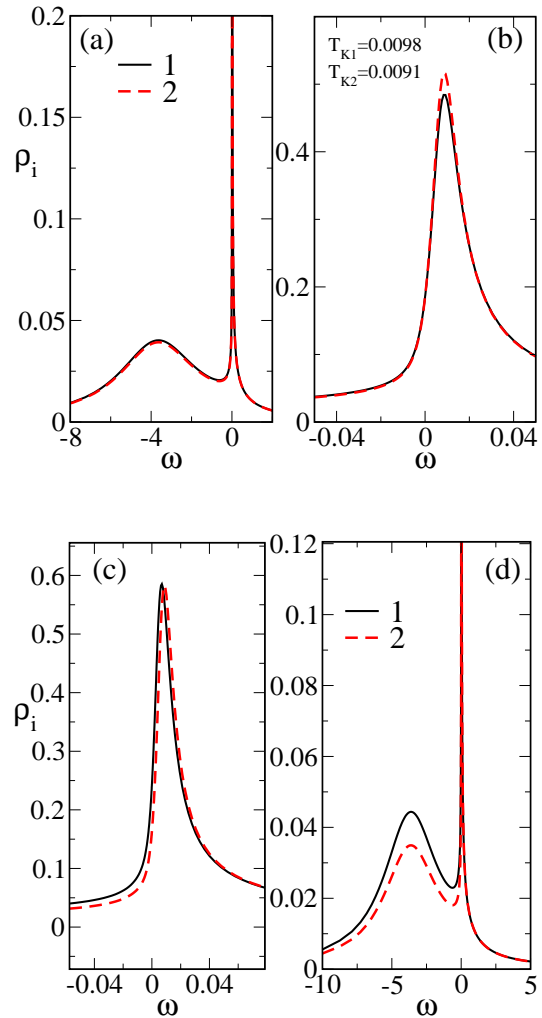

\begin{center}
\vspace{0.8cm}
\includegraphics[width=7cm]{compara_na.eps} \\
\vspace{0.8cm}
\includegraphics[width=7cm]{compara_nb.eps}
\caption{(color online) Spectral density of dot 1 (black) and dot 2 (red) 
as a function of energy for  $\Gamma_1=1$, $\Gamma_2=0.9$, $E_1=-4$ and (a), (b): 
$\delta=-0.0305$ and $T=0.37\ 10^{-2}$, (c), (d): $\delta=-0.029$ and 
$T=0.25\ 10^{-2}$.}
\label{Fig:compara2}
\end{center}
\end{figure}

Depending on the particular property that is studied, the tuning of the parameters is slightly different to get 
the effective SU(4) symmetry for this property. This is illustrated in  Fig. \ref{Fig:compara2}, where the 
densities of states are compared for two conditions different from $n_1=n_2$ discussed above. In 
Fig. \ref{Fig:compara2} top, the parameters are tuned in such a way that the densities coincide at the Fermi 
level: $\rho_1(0)=\rho_2(0)$. This condition render the densities of states very similar in the whole energy range. 
The occupations are slightly different, 
$n_1=0.470$ and $n_2=0.465$, signaling a deviation from the Friedel sum rule, Eq. (\ref{fsr}) valid at $T=0$.
The maximum of the Kondo resonance of the doublet 2 (that with lowest energy) is higher, although both maxima 
lie nearly at the same position, and the shape of the resonance corresponds to the SU(4) Kondo effect. 
However the half width at half maximum are slightly different: $T_{K1}=0.0098$ and $T_{K2}=0.0091$
for doublets 1 and 2 respectively.

In Fig. \ref{Fig:compara2} bottom, the parameters are chosen to get the same value of the maximum of the Kondo resonance 
$\rho_1^{max}=\rho_2^{max}$. When this condition is satisfied, the weight of the charge transfer peak 
(and the corresponding occupation $n_i$) for each doublet differ, being larger for dot 1.
The resonance in $\rho_2$ is displaced slightly to the right with respect to $\rho_1$. However,
both densities of states near the Fermi level are very similar. This implies that in suitable conductance 
experiments, the conductance through both dots $G_i(V)$ are proportional. These experiments 
correspond to asymmetric arrangements such that the coupling to source and drain leads differ by a factor of the
order of 10 or larger, and $G_i(V)=dI_i/dV_i$ is measured, where $I_i$ is the current through dot $i$ and 
$V_i$ is the voltage of the lead (source or drain) less coupled to dot $i$.\cite{oks} This is a situation similar 
to that in scanning-tunneling-spectroscopy experiments. In the conditions of Fig. \ref{Fig:compara2} bottom,
$G_1(V)/G_2(V)=A_1/A_2$, where the constant asymmetry factors are 
$A_i=4 \Gamma_{Si} \Gamma_{Di}/(\Gamma_{Si}+\Gamma_{Di})^2$.

\section{Summary and discussion}
\label{summ}

We have considered an Anderson model that describes two capacitively coupled
quantum dots, each one connected to a drain and a source lead. We have investigated the possibility that 
the SU(4) symmetry, lost at the Hamiltonian level when the total coupling to the leads
are different ($\Gamma_1 \neq \Gamma_2$), can be restored at low energies as an emergent symmetry,\cite{eme}
by changing the difference of on-site energies $\delta=E_2-E_1$.
We find that for small temperatures (specifically lower that the Kondo temperature
of the SU(4) Kondo effect $T_K^{\rm SU(4)}$), it is possible to tune $\delta$
such that the Kondo resonances for each dot sensed by suitable chosen 
conductance experiments are proportional. Specifically at this value of $\delta$,
the conductance through
each dot $G_i(V)$ in a configuration of voltages similar to those used in 
scanning-tunneling-spectroscopy experiments, have the same line shape within
experimental errors and reflect the characteristic shape of the SU(4) Kondo 
resonance in the spectral density. However, for this value of $\delta$,
the total occupations $n_i$ for each dot are slightly different, pointing out
the absence of full SU(4) symmetry at large energies.
 
The temperature dependence of the conductances at not too high temperatures, 
also corresponds to the SU(4) regime rather that the SU(2) one, 
since it is given by the energy and temperature dependence of the densities $\rho_i$.
However, slight differences in the line shape of both conductances 
can appear as a function of temperature, because both spectral densities and occupations
do not have exactly the same temperature dependence.

From the theoretical point of view it remains to study more accurately 
with alternative techniques, to what extent the SU(4) symmetry is kept at
the lowest energies. The NCA is not reliable at temperatures well below 
the Kondo one and the Friedel sum rule is not reproduced 
with an error of about
15 \% at very low temperatures. Combining NRG and RPT the low-energy Fermi-liquid 
properties might be studied in detail. A difficulty for numerical studies is the fine 
tuning in $\delta$ required to obtain a manifestation of SU(4) symmetry
in a given property.

\section*{Acknowledgments}

We are partially supported by CONICET,
Argentina. This work was sponsored by PICT 2010-1060 and 2013-1045 of the
ANPCyT-Argentina and PIP 112-201101-00832 of CONICET.

\end{document}